\documentclass[11pt,preprintnumbers,amsmath,amssymb]{revtex4}
\usepackage{graphicx}
\usepackage{epsfig}
\usepackage{bm}

\newcommand{\p}{\phi}

\begin{document}

\title{Cyclic Universes from General Collisionless Braneworld Models}

\author{E.~N.~Saridakis
\footnote{E-mail: msaridak@phys.uoa.gr}} \affiliation{Department
of Physics, University of Athens, GR-15771 Athens, Greece}

\begin{abstract}
We investigate the full $5D$ dynamics of general braneworld
models. Without making any further assumptions we show that cyclic
behavior can arise naturally in a fraction of physically accepted
solutions. The model does not require brane collisions, which in
the stationary case remain fixed, and cyclicity takes place on the
branes. We indicate that the cosmological constants play the
central role for the realization of cyclic solutions and we show
that its extremely small value on the observable universe makes
the period of the cycles and the maximum scale factor
astronomically large.
\end{abstract}

 \maketitle

\section{Introduction}

The last decade proves to be really exciting for cosmology.
Observational data indicated, among other very interesting
results, that the expansion of the universe is accelerated
\cite{observ}. At the same time the braneworld scenario appeared
in the literature \cite{Horawa,RS99}. Though the exciting idea
that we live in a fundamentally higher-dimensional spacetime which
is greatly curved by vacuum energy was older \cite{Rubakov83}, the
new class of ``warped" geometries offered a simple way of
localizing the low energy gravitons on the brane.

In this novel background the old idea of a cyclic Universe was
reheated. Started as ekpyrotic \cite{Khoury.best,Khoury.rebound},
enriched to ekpyrotic/cyclic
\cite{ekpyrotic1,Khoury.rebound,Turok.simplified,Turok.sing,ekpyrotic.pert,cyclic.clifton,chargeBH}
and recently to new ekpyrotic
\cite{ekpyrotic3,ekpyrotic4,ekpyrotic2,ekpyrotic.fields}, the new
paradigm tries to be established as an alternative to standard
cosmology. According to its basic contents, our universe
experiences an infinite or extremely large number of cycles, each
one consisting of a hot bang phase, a phase of accelerated
expansion, a phase of slow-ekpyrotic contraction and a bounce-bang
that triggers the next cycle. Starting with a simplified notional
framework (infinite and not ``created" time) cyclic cosmology have
many advantages. It successfully faces the homogeneity, isotropy,
topological and flatness problems, it handles the issue of initial
conditions, it incorporates the dark energy and transforms it to
an important factor, and it provides the mechanism of the
generation of cosmic perturbations and of structure formation.
However, there are some key issues that do not have a consistent
and efficient approach so far, despite the great progress. These
are the settlement of the singularity, although temperature and
density remain finite, the entropy evolution, and the fate of the
perturbations through the bounce. Through this research, cyclic
scenarios have become more complicated, by the insertion of more
complex potentials, of more branes \cite{Khoury.best}, of the
mechanism of ghost condensation \cite{ghcond,ekpyrotic3}, of more
scalar fields \cite{ekpyrotic.fields} and of procedures which
cancel the tachyonic instabilities \cite{ekpyrotic4}.

Most of the works on cyclic cosmology involve, initially or at
some stage, the transition to effective $4D$ equations. However,
as it was mentioned in \cite{Linde01,4Dbreakdown}, such a
procedure does not lead to reliable results since one cannot
return to the $5D$ description self-consistently. Furthermore, the
old $4D$-singularity problem (of both Big Bang and traditional
cyclic universes), has been replaced by a new one (equally
annoying) concerning the singularity of extra dimension(s). This
later case is accompanied by the brane collision phenomenon, which
seems to be a basic constituent of the ekpyrotic scenario.

In this work we desire to investigate the full $5D$ dynamics of
general braneworld models and examine if a cyclic behavior is
possible. This is an essential procedure in order to consistently
confront the arguments of the authors of \cite{Linde01}, which
claim that cyclic behavior cannot arise from a complete $5D$
description, and our study must not include any additional
assumptions or fine tunings in order to remain general and
therefore convincing. Secondly, we are interested to explore if a
cyclic behavior of $5D$ dynamics is necessarily related to brane
collisions. This work is organized as follows: In section
\ref{model} we present the $5D$ braneworld model and we derive the
equations of motion. In section \ref{analyt} we provide analytical
solutions for two simplified stationary solution subclasses, while
in \ref{numer} we investigate numerically the full stationary
dynamics. Finally, in section \ref{discussion} we discuss the
physical implications of our analysis and we summarize the
obtained results.

\section{The model}\label{model}

We consider quite general braneworld models, characterized by the
action \cite{brcod,sarjcap}:
\begin{equation}
\kappa^2_5 S=\frac{1}{2}\int d^4xdy\sqrt{-g}\,R+\int
d^4xdy\sqrt{-g}\left[-\frac{1}{2}(\partial\p)^2-V(\p)\right]-\sum_{i=1,2}\int_{b_i}d^4x\sqrt{-\gamma}\,\{[K]+U_i(\p)\},
\label{action}
\end{equation}
where $\kappa^2_5=\frac{1}{M^3_5}$ is a $5D$ gravitational
constant, and all quantities are measured in units of $M_5$. The
first term describes gravity in the five dimensional bulk space.
The second term corresponds to a minimally coupled bulk scalar
field with the potential $V(\p)$. The last term corresponds to two
$3+1$ dimensional branes, which constitutes the boundary of the
$5D$ space. We allow for a potential term $U(\p)$ for the scalar
field at each of the two branes, and we denote by $\gamma$ the
induced metric on them and by $K$ their extrinsic curvature. Here
and in the following the square brackets denote the jump of any
quantity across a brane ($[Q]\equiv Q(y_{+})-Q(y_{-})$). The
reason we use a second brane is to eliminate possible ``naked''
singularities, therefore by assuming $S^1/{\mathbb Z}_2$ symmetry
across each brane we restrict our interest only in the interbrane
space.

As usual the two branes are taken parallel, $y$ denotes the
coordinate transverse to them and we assume isometry along three
dimensional $\mathbf{x}$ slices including the branes. For the
metric we choose the conformal gauge \cite{brcod,sarjcap}:
\begin{equation}
ds^2=e^{2B(t,y)}\left(-dt^2+dy^2\right)+e^{2A(t,y)}d\mathbf{x}^2.
\label{metric}
\end{equation}
This metric choice, along with the residual gauge freedom
$(t,y)\rightarrow(t',y')$ which preserves the $2D$ conformal form,
allows us to ``fix" the positions of the branes. Without loss of
generality we can locate them at $y=0,1$, having in mind that
their physical distance is encoded in the metric component
$B(t,y)$, and at a specific time it is given by
\cite{brcod,sarjcap}:
\begin{equation}
D(t)\equiv\int_0^1dy\sqrt{g_{55}}=\int_0^1dy\,e^{B(t,y)},
\label{distance}
\end{equation}
quantity that is invariant under the residual gauge freedom in our
coordinates. The reason we prefer the metric (\ref{metric}),
instead of  the usual form in the literature, is that in the later
case the brane positions are in general time-dependent  and the
various boundary conditions are significantly more complicated.
Thus, our coordinates are preferable for numerical calculations,
despite the loss of simplicity in the definitions of some
quantities. Eventually, the physical interpretation of the results
is independent of the coordinate choice.

The non-trivial five-dimensional Einstein equations consist of
three dynamical:
\begin{eqnarray}\
\ddot{A}-A''+3\dot{A}^2-3A'^2=\frac{2}{3}e^{2B}V\nonumber\\
\ddot{B}-B''-3\dot{A}^2+3A'^2=-\frac{\dot{\p}^2}{2}+\frac{\p'^2}{2}-\frac{1}{3}e^{2B}V\nonumber\\
\ddot{\p}-\p''+3\dot{A}\dot{\p}-3A'\p'+e^{2B}V_{,\p}=0,
\label{eomfull}
\end{eqnarray}
and two constraint equations:
\begin{eqnarray}\
-A'\dot{A}+B'\dot{A}+A'\dot{B}-\dot{A}'=\frac{1}{3}\dot{\p}\p'\nonumber\\
2A'^2-A'B'+A''-\dot{A}^2-\dot{A}\dot{B}=-\frac{\dot{\p}^2}{6}-\frac{\p'^2}{6}-\frac{1}{3}e^{2B}V\,,
\label{constraints}
\end{eqnarray}
where primes and dots denote derivatives with respect to $y$ and
$t$ respectively. It is easy to show that the constraints are
preserved by the dynamical equations.

Additionally, from the boundary terms in the action for the branes
we obtain the following junctions (Israel) conditions:
\begin{eqnarray}\
[A']=\mp\frac{1}{3}\,Ue^B\nonumber\\
\,[B']=\mp\frac{1}{3}\,Ue^B\nonumber\\
\,[\p']=\pm e^B\,U_{,\p}, \label{junctions}
\end{eqnarray}
where the upper and lower signs refer to the branes at $y=0,1$,
respectively. Since we have imposed ${\mathbb Z}_2$ symmetry
across the two branes, for any function $Q$ we get:
\begin{equation}
[Q']_0=2Q'(0^{+})\ \ ,\ \ \ \ \    [Q']_1=-2Q'(1^{-}). \label{z2}
\end{equation}

For the bulk potential $V(\p)$ we assume a general form consistent
with the stabilization mechanism \cite{Goldberger99}:
\begin{equation}
V(\p)=\frac{1}{2}m^2\p^2+\Lambda \label{bulkpot},
\end{equation}
where $\Lambda$ is the $5D$ bulk cosmological constant. For the
brane potentials $U_i(\p)$ we use the quite general quadratic form
\cite{brcod,sarjcap}:
\begin{equation}
U_i(\p)=\frac{1}{2}M_i(\p_i-\sigma_i)^2+\lambda_i
\label{branepot},
\end{equation}
where $\lambda_i$ stand for the brane tensions and $\p_i$ for the
value of $\p$ on the i-th brane. The ``masses" $m$ and $M_i$ can
be varying.

Finally, the induced $4D$ metrics of the two (``fixed''-position)
branes in the conformal gauge are simply given by
\begin{equation}
ds^2=-d\tau^2+a^2(\tau)\,d\mathbf{x}^2, \label{4Dmetric}
\end{equation}
with $d\tau_i=e^{B_i}dt$ and $a_i=e^{A_i}$ the proper times and
scale factors of the two branes. Thus, for the Hubble parameter on
the branes we acquire
\begin{equation}
H_i\equiv\frac{1}{a}\frac{da}{d\tau}\Big\vert_i
=e^{-B_i}\dot{A}_i, \label{Hubble}
\end{equation}
which is invariant under residual gauge transformations. As usual,
we identify the brane at $y=0$  as the visible brane corresponding
to our Universe. Note that relations (\ref{4Dmetric}) and
(\ref{Hubble}) can be generalized to hold in every 3+1 slice
transverse to the fifth dimension.

Finally, let as comment on the behavior of gravity on the physical
brane in our model. As was shown in \cite{grav1}, in such a
two-brane model we re-obtain the correct Newton's law in the
brane-universe. This becomes more transparent if we include the
aforementioned brane stabilization mechanism \cite{Goldberger99},
where the two-brane model leads naturally to the recovery of the
Einstein gravity on the physical brane \cite{brcod,grav2}. More
generally, even in a generic two-brane model with arbitrary bulk
and brane potentials, using the renormalization group flow of the
4D Newton's constant in IR and UV, it can be shown that Newton's
law is recovered, plus one extremely small brane correction
\cite{grav3}.

\section{Stationary case. analytical solutions}\label{analyt}

The model we have described is general enough and includes the
full spacetime evolution of the $5D$ braneworld. Our aim is to
look for physically accepted solutions that correspond to cyclic
behavior. The main difficulty of solving the equation system
(\ref{eomfull})-(\ref{branepot}) is that one has to satisfy the
constraints (\ref{constraints}) and boundary conditions
(\ref{junctions}) at $t=t_0$, ensuring that no divergencies are
present in the interbrane space. Then, one has to assure that time
evolution will not be unstable, or give rise to naked
singularities in the bulk.

Let us first investigate a subclass of solutions, the so-called
stationary ones. In this case we assume that
\begin{eqnarray}
B(t,y)\rightarrow B(y)\ \ \ \ \ \ \ \ \ \nonumber\\
A(t,y)\rightarrow B(y)+H\,t. \label{stationary}
\end{eqnarray}
This case is characterized by a fixed bulk geometry and maximally
symmetric (de Sitter, Minkowski or oscillatory) branes:
\begin{equation}
ds^2=e^{2B(y)}\left(dy^2-dt^2+e^{2Ht}d\mathbf{x}^2\right).
\label{metricstat}
\end{equation}
Note that there is a name confusion in the literature, since
spacetimes (\ref{stationary}) are called stationary in
\cite{Tyewass01c} and when $H=0$ they are called static (a
convention that we follow in this paper), while in \cite{brcod}
they are called static even for $H\neq0$.

 Under (\ref{stationary}) the system of equations (\ref{eomfull}) and (\ref{constraints})
 transits to the time-independent form:
\begin{eqnarray}
B''(y)=-\frac{1}{6} e^{2 B(y)}-\frac{\p'^2(y)}{4}\nonumber\\
\p''(y)+3 B'(y)\p'(y)-e^{2 B(y)}\frac{\partial{V(\p)}}{\partial\p}=0\nonumber\\
H^2=B'^2(y)+\frac{1}{6} e^{2 B(y)}V(\p)-\frac{\p'^2(y)}{12},
\label{eomstat}
\end{eqnarray}
where $\p$ is also assumed to be time independent. Furthermore,
boundary conditions (\ref{junctions}) will be satisfied at all
times, provided that they do so at $t=t_0$. Finally, according to
definition (\ref{Hubble}) and under the ansatz (\ref{stationary}),
the Hubble parameter on the physical brane is simply
$H_0^2=e^{-2B(0)}H^2$.

Equations (\ref{eomstat}) and boundary conditions
(\ref{junctions}) cannot be solved analytically for the general
bulk and brane potentials of (\ref{bulkpot}) and (\ref{branepot}).
However, this is possible in two simplified cases: A) Assuming
$V(\p)=0$ with full $U_i(\p)$, and B) assuming $V(\p)=\Lambda$,
$U_0(\p)=\lambda_0$ and $U_1(\p)=\lambda_1$. In the following
subsections we examine these two cases successively.

\subsection{$V(\p)=0$ and
$U_i(\p)=\frac{1}{2}M_i(\p_i-\sigma_i)^2+\lambda_i$}\label{A}

If we set $V(\p)=0$ we can acquire analytical solutions depending
on the sign of $H^2$. For $H^2>0$ and setting $H=|\sqrt{H^2}|$ we
acquire:
\begin{equation}
B(y)=B(0)+\frac{1}{3} \log{\left(\frac{B'(0)}{H}
\sinh{3Hy}+\cosh{3Hy}\right)} \label{Byp}
\end{equation}
 and
\begin{equation}
\p(y)=\p(0)-\frac{2}{\sqrt{3}}\log{\left[\frac{\left(e^{3Hy}-u\right)\left(1+u\right)}
{\left(e^{3Hy}+u\right)\left(1-u\right)}\right]} \label{pyp}
\end{equation}
where we set $u=\sqrt{\frac{B'(0)-H}{B'(0)+H}}$, and $|B'(0)|>H$
always in this case if we want $\p'(0)$ to be real.

For $H^2<0$ and setting $\theta=|\sqrt{-H^2}|$ we get:
\begin{equation}
B(y)=B(0)+\frac{1}{3} \log{\left(\frac{B'(0)}{\theta} \sin{3\theta
y}+\cos{3\theta y}\right)} \label{Byn}
\end{equation}
and
\begin{equation}
\p(y)=\p(0)-\frac{2}{\sqrt{3}}\log{\left[\frac{\tan\left(\frac{3}{2}\theta
y+\frac{1}{2}\arctan{\frac{\theta}{B'(0)}}\right)}
{\tan\left(\frac{1}{2}\arctan{\frac{\theta}{B'(0)}}\right)}\right]}.
\label{pyn}
\end{equation}

Let us make some comments here. Firstly, in the expressions above
we have eliminated the three integration constants in terms of
$B(0)$, $B'(0)$ and $\p(0)$, since it is more convenient, but one
could equally use the values of three of $B$, $B'$, $\p$ and $\p'$
at any $y$. Moreover, these solutions have to satisfy the boundary
conditions (\ref{junctions}). Eventually, all quantities,
including $H^2$, are given in terms of the six parameters of the
model ($M_0,\lambda_0,\sigma_0,M_1,\lambda_1,\sigma_1$, since we
have set $m=0$ and $\Lambda=0$). We will return to this subject in
the next section. Secondly, note that there is analytical
continuation between the solutions for positive and negative
$H^2$, i.e. the two solutions coincide for $H\rightarrow i
\theta$. Thirdly, it is easy to see that in the limit
$H^2\rightarrow 0$, $B(y)$ and $\p(y)$ become smoothly a negative
logarithm of a linear in $y$ function. Lastly, note that due to
the non-linear form of the differential equations there exist more
solution branches which can be obtained straightforwardly from
(\ref{Byp})-(\ref{pyn}) by sign changing. However, since all of
them imply the same metric for the visible brane, which is given
below and which is the central subject of this work, we do not
write them explicitly.

In order for the solutions to be physically accepted, we have to
assure that no naked singularities are present in the interbrane
space, i.e. between $y=0$ and $y=1$. The investigation of the
$H^2>0$ case was presented in \cite{Manos.Braneworld} where we
showed that $H^2$ values are restricted to very small values and
only one $5D$ sub-surface of the $6D$ parameter space allows for
arbitrary large $H^2$ values. Here we repeat the cogitations in
the $H^2<0$ case. Solutions (\ref{Byn}) and (\ref{pyn}) have poles
at $\frac{B'(0)}{\theta} \sin{3\theta y_p}+\cos{3\theta y_p}=0$,
i.e. at
\begin{equation}
y_{pn}=\frac{1}{3 \theta}\,\left(y_{0}+n\pi\right)
\end{equation}
with $n\in {\mathbb Z}$, where
$y_{0}=\arctan\left(-\frac{\theta}{B'(0)}\right)$ with $-\pi\leq
y_{0}<0$. In this notation $y_{p0}$ stands for the largest
negative pole.
 Demanding none of
these $y_{pn}$ lying in [0,1] we require that the smallest
positive pole, i.e. $y_{p1}$, to be greater than 1. This leads to
$y_0>3\theta-\pi$ and since $-\pi\leq y_{0}<0$ we acquire:
\begin{equation}
3\theta-\pi<\arctan\left(-\frac{\theta}{B'(0)}\right)<0.\label{wind2}
\end{equation}
Thus, this relation provides a narrow and absolute window for
$\theta$, in a sense that there are no areas at all in the $6D$
parameter space that give $\theta$ larger than $\pi/3$.

Finally, we desire to explore the forms of the bulk and brane
geometries in these cases. For $H^2\geq0$ we get de Sitter,
anti-de Sitter or Minkowski branes (3+1 slices in general), that
have been studied extensively in the literature. The bulk
structure is fixed and this can be also confirmed by calculating
the brane physical distance according to (\ref{distance}), which
is obviously time independent (given analytically as an expression
in terms of hypergeometric and hyperbolic functions of the
parameters), i.e. the branes do not move.

For $H^2<0$, which is the case of interest in this work, we
acquire also a fixed bulk with constant brane distance. For the
branes we get an interesting oscillatory behavior. In particular,
the $5D$ metric: $ds^2=e^{2B(y)}\left(-dt^2+dy^2+e^{2i\theta
t}d\mathbf{x}^2\right)$, implies a $4D$ metric of the form
(\ref{4Dmetric}) on the visible brane, with the scale factor given
as:
\begin{equation}
a^2(\tau)=c\,\sin^2\left(e^{-B(0)}\theta \tau\right),
 \label{osc1}
\end{equation}
where $\tau$ is the visible brane proper time defined above as
$d\tau=e^{B(0)}dt$ (for clarity in the following we omit the index
$i$ which distinguishes the proper times of the two branes, using
$\tau$ for the visible brane and $\tau_1$ for the hidden one). The
constant $c$ depends on the exact relation between $\tau$ and bulk
time $t$ (it is equal to $e^{2B(0)}$ for $\tau=e^{B(0)}t$).
Moreover, one should add a constant phase which would be
determined by the $a^2(\tau)$ value at a specific $\tau$. Without
loss of generality we have performed a shift in $\tau$ such that
$a^2(0)=0$ (note that $\tau=0$ is not an initial time but just a
random one). Definitely, one could achieve a relation similar to
(\ref{osc1}) in terms of bulk time $t$, but we desire to present
the results from the visible brane observer's point of view.

\subsection{$V(\p)=\Lambda$,
$U_0(\p)=\lambda_0$ and $U_1(\p)=\lambda_1$}\label{B}

In the case where the branes have just tensions, the bulk
potential is comprised of a non-zero cosmological constant and
there is no scalar field, analytical solutions, depending on the
sign of $H^2$, can be derived.

For $H^2>0$, with $H=|\sqrt{H^2}|$ we acquire:
\begin{equation}
B(y)=-\log{\left(\frac{c_1 e^{H y}}{2H^2}+\frac{\Lambda e^{-H
y}}{12 c_1}\right)},\label{BypB}
\end{equation}
where
$c_1=H^2\left[e^{-B(0)}-\sqrt{e^{-2B(0)}-\Lambda/(6H^2)}\right]$,
similar to the results of \cite{analsol}. Expression (\ref{BypB}),
apart from $y\rightarrow\pm\infty$, has a naked singularity at
$y_p=\frac{1}{2H}\log{(-\frac{H^2\Lambda}{6c_1^2})}$ which must
lie outside $[0,1]$. It is obvious that when $\Lambda<0$ it is
singularity-free and is the well studied AdS bulk case of the
literature. Furthermore, solution (\ref{BypB}), in the specific
case where $\Lambda<0$, $\lambda_0=\sqrt{-6\Lambda}$ and
$\lambda_1=-\sqrt{-6\Lambda}$, corresponds to the two-brane
Randall-Sundrum model \cite{RS99}, where $H^2$ acquires a zero
value and $B(y)$ becomes a negative logarithm of a term linear in
$y$.

For $H^2<0$, and setting $\theta=|\sqrt{-H^2}|$ we get:
\begin{equation}
B(y)=\log{\left[\frac{\theta}{\sqrt{-\frac{\Lambda}{6}}}\,\frac{1}{\sin{\left(\theta
y+\arcsin{\frac{\theta
e^{-B(0)}}{\sqrt{-\frac{\Lambda}{6}}}}\right)}}
\right]}.\label{BynB}
\end{equation}
Note that in this case $\Lambda$ must be negative, as it is easily
implied from the last of equations (\ref{eomstat}) if we demand
$B(y)$ to be real, i.e. the bulk is always AdS. Expression
(\ref{BynB}) possesses singularities at
\begin{equation}
y_{pn}=\frac{1}{\theta}\,\left(y_{0}+n\pi\right)
\end{equation}
with $n\in {\mathbb Z}$, where $y_{0}=-\arcsin\left(\frac{\theta
e^{-B(0)}}{\sqrt{-\frac{\Lambda}{6}}}\right)$ with $-\pi\leq
y_{0}<0$. As previously, $y_{p0}$ stands for the largest negative
pole. Forcing the smallest positive pole, i.e. $y_{p1}$, to be
greater than 1 we acquire $y_0>\theta-\pi$, and since $-\pi\leq
y_{0}<0$ we finally obtain:
\begin{equation}
\theta-\pi<-\arcsin\left(\frac{\theta
e^{-B(0)}}{\sqrt{-\frac{\Lambda}{6}}}\right)<0.\label{wind2b}
\end{equation}
Thus, in this case there are no areas at all in the $3D$ parameter
space that lead to solutions with $\theta$ larger than $\pi$.

Let as describe the spacetime geometry that corresponds to these
solutions (we repeat that there are more solution branches arising
form (\ref{BypB}) and (\ref{BynB}) by sign changing, but they
correspond to the same  brane metric). In the $H^2\geq0$ case we
acquire two stabilized branes, with their time-independent
physical distance given analytically through (\ref{distance}) in
terms of inverse hyperbolic trigonometric functions of the
parameters. The $4D$ induced geometry on the two branes can be de
Sitter, anti-de Sitter or Minkowski and has been investigated
widely in the literature. For $H^2<0$ we also obtain a fixed bulk
with the time-independent physical distance of the two branes
given easily analytically. However, the geometry on the branes
acquires the interesting oscillatory behavior described in the
previous subsection, with a scale factor of the form:
\begin{equation}
a^2(\tau)=c\,\sin^2\left(e^{-B(0)}\theta \tau\right),
 \label{osc2}
\end{equation}
identically to relation (\ref{osc1}).

Relations (\ref{osc1}) and (\ref{osc2}) correspond to cyclic
Universes. Indeed, they imply that the scale factor pulsates
between the following extremum values:
\begin{eqnarray}
a_{min}=0,\nonumber\\
 a_{max}=\sqrt{c},
 \label{minmax}
\end{eqnarray}
where as we mentioned $c$ depends on the specific relation between
$t$ and $\tau$ ($c=e^{2B(0)}$ for $\tau=e^{B(0)}t$). The constant
period of the cycles is given as:
\begin{equation}
T=\frac{\pi e^{B(0)}}{\theta}.
 \label{period}
\end{equation}
In (\ref{minmax}),(\ref{period}) all quantities are measured in
units of $M_5$. The aforementioned behavior corresponds to a
sequence of expanding and contracting phases on the branes, while
the physical brane positions stay constant. The fifth dimension
remains unaffected, that is there are no brane collisions. The
bulk dynamics determines also the quantitative characteristics of
the cycles. Indeed, the value of $B(y)$ on the physical brane
specifies the maximum value of the scale factor and can be
arbitrary. Moreover, it designates the cycles period. Fortunately,
the fact that $\theta$ is bounded from above (according to
(\ref{wind2}) and (\ref{wind2b})) restricts completely its effect
on decreasing the period. We will return to these subjects in the
discussion section below, where we show that only astronomically
large maximum scale factors and periods are possible. Finally,
note that in the aforementioned analysis cyclicity arises
naturally from the $5D$ dynamics, and the oscillatory behavior of
the scale factor is explicit and not a result of Hubble constant
sign change in special cases
\cite{Turok.simplified,ekpyrotic3,cyclic.clifton,holographic}.
Furthermore, it emerges from general braneworld models, without
Gauss-Bonnet terms \cite{TetradGB} or the assumption of charged
AdS bulk black holes which charge (along with fine tuning) is
responsible for restricted cyclic behavior \cite{chargeBH}.

We close this section by referring to the stability of the
aforementioned solutions in both metric and scalar field
components. As it was shown in \cite{Tyewass01c,stabil}, a
two-brane model with the bulk and brane potentials considered
above, has stable solutions, free of tachyonic modes. In
particular, the presence of the  stabilization mechanism
\cite{Goldberger99}, which forbids branes to move, makes all
physically accepted solutions (i.e. those which are
singularity-free and satisfy boundary conditions) to be stable
 under perturbations \cite{grav2}. This feature has been also
confirmed by numerical investigation in \cite{brcod} and has been
verified by us, too. Thus, the stability of the solutions provides
the necessary physical hypostasis to our model.

\section{Stationary case. Numerical results}\label{numer}

In the previous section we derived analytical solutions for the
full $5D$ equations, in the stationary ansatz with two simplified
potential cases. We expressed the solutions in terms of the values
of $B(0)$, $B'(0)$ and $\p(0)$ and we stated that these must
satisfy the boundary conditions (\ref{junctions}). The usual, in
the literature, method to achieve this is to randomly choose $H^2$
and the solution values at $y=0$ and then fine-tune the model
parameters. However, this procedure definitely does not reveal the
properties and the rich structure of the solutions. On the
contrary, as we showed in \cite{Manos.Braneworld}, it restricts
the investigation in a small subclass of solutions. The natural
way to encounter the problem is to choose randomly and uniformly
the potential parameters at first, and then seek for physically
accepted solutions, i.e. divergencies-free expressions which
satisfy equations (\ref{eomstat}) and boundary conditions
(\ref{junctions}). This is a hard task in general. The method we
use is the following: We first choose randomly the values of the
model parameters, uniformly distributed in a hyper-cube ($6D$ in
the case of subsection \ref{A} and 3D in that of \ref{B}). The
obtained results do not depend on the hyper-cube's size, but on
its effectual covering (number of parameter multiplets used in the
calculation). We use the globally convergent Schmelcher-Diakonos
algorithm \cite{Diakonos} to solve the transcendental equation
system with accuracy $10^{-13}$. We find hexads (or triads) of
parameters corresponding to acceptable solutions, and calculate
$H_0^2$ through $H_0^2=e^{-2B(0)}H^2$.

Numerical investigation reveals that only a small fraction of
parameter choices (we use $10^6$ parameter multipltes) allows for
solutions to exist ($\approx15\%$ in case \ref{A} and $\approx
5\%$ in case \ref{B}). Note that, as we mentioned in
\cite{Manos.Braneworld}, the parameter sub-space that leads to
solutions is neither compact nor uniform. The question if it
consists from continuous areas or from independent points does not
have a clear answer \cite{Tetradis01}. Now, inside these
percentages only a small fraction corresponds to solutions with
$H^2<0$, i.e. to oscillatory ones ($\approx2\%$ in case \ref{A}
and $\approx 6\%$ in case \ref{B}). The reason of the
significantly smaller appearance of $H^2<0$ solutions is that they
posses many singularities, in comparison with the $H^2>0$ case,
and therefore it is harder to find a solution with no singularity
at $[0,1]$. In conclusion, in total only about $\approx10^{-1}\%$
of the random and uniform parameter choices correspond to
oscillatory behavior of the metric of the visible brane. In these
cases, (\ref{osc1}) and (\ref{osc2}) are numerically verified and
the analysis of the cyclic solution is valid.

\begin{figure}[h]
\begin{center}
\mbox{\epsfig{figure=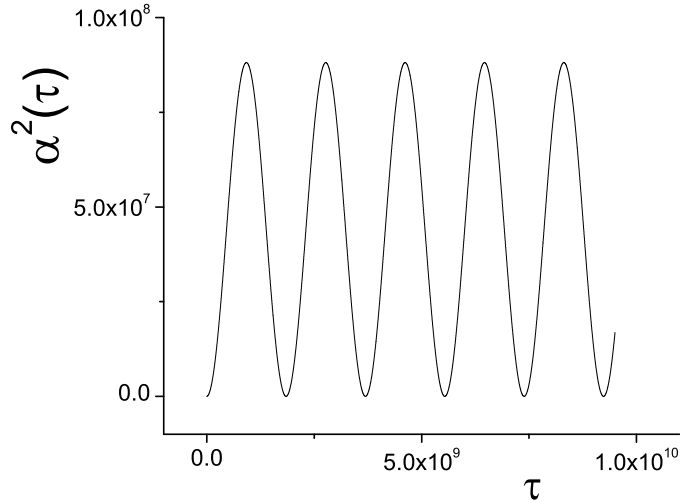,width=9cm,angle=0}} \caption{\it
Visible brane scale factor evolution for one random solution of
the stationary case with the general bulk and brane potentials of
(\ref{bulkpot}) and (\ref{branepot}). The edge of the parameter
space hyper-cube extends from -$10^3$ to $10^3$. The solution
arises from $m=81.03$, $\Lambda=-505.8$, $\lambda_0=-212.5$,
$\lambda_1=-464.2$, $M_0=77.94$, $M_1=180.7$, $\sigma_0=1.308$,
$\sigma_1=-3.556$ (for simplicity we provide the first four
relevant digits only). The obtained $B(0)$ and $\theta$ values are
correspondingly $9.148$ and $0.1522$, and $\tau$ is calculated as
$e^{B(0)}t$. All quantities are expressed in $M_5$ units.}
\label{fig}
\end{center}
\end{figure}

Since we have confirmed our analytical calculations numerically,
we can proceed to the numerical investigation of the general
stationary case, i.e. with the full (\ref{bulkpot}) and
(\ref{branepot}) potentials, where analytical solutions cannot be
obtained. Again, we randomly choose the values of the 8 model
parameters ($10^6$ octads of
$M_0,\lambda_0,\sigma_0,M_1,\lambda_1,\sigma_1$, $m$ and
$\Lambda$), from a uniform distribution, and we solve the
transcendental equation system of (\ref{eomstat}) and
(\ref{junctions}). Note that the parameter space hyper-cube is
taken large and symmetric (its edge extends from -$10^3$ to
$10^3$) and we do not impose any constraints on the parameter
values (assuming for example positive ``masses" or opposite
tension branes) in order to remain as general as possible. In this
case only $\approx 0.6\%$ of the parameter multiplets corresponds
to physically accepted solutions, and within them only $\approx
4\%$ corresponds to $H^2<0$, i.e. in total only $\approx
10^{-2}\%$ of the parameter choices lead to cyclic branes. We
mention that these percentages increase radically if we restrict
the investigation in specific parameter signs. In fig.~\ref{fig}
we present the evolution of the scale factor of the visible brane
for one such solution.
We conclude that even in the absence of analytical calculations in
the general-potential stationary case, we do obtain fixed and
cyclic 3+1 branes and relations (\ref{osc1}) (or (\ref{osc2})) and
(\ref{minmax}),(\ref{period}) are satisfied. In other words, if
the requirements for a complete $5D$ solution to exist are
fulfilled, then (\ref{osc1}) (or (\ref{osc2})) and
(\ref{minmax}),(\ref{period}) are valid analytically, with the
numerics only determining $B(0)$ and $\theta$. The only numerical
restriction arises from the presence of $B(0)$ in an exponential,
which prevents us from handling arbitrary large values as in the
analytical calculations. However, this can be solved by the
additional scaling transformation proposed in \cite{brcod}.

Let us make a comment here concerning the ``singular'' points
where the scale factor vanishes. These points correspond to the
so-called ``bounce'', which is always present in all
cyclic-cosmology models. Although there have been many attempts in
the literature to avoid the involved singularity, none of them is
completely satisfactory up to now. These approaches, such is the
insertion of quantum fluctuations
\cite{Bojowald01,ekpyrotic3,ekpyrotic4} or the use of
 loop quantum gravity  modifications \cite{Maartens},
 could be included in our analysis, leading to a smoothing out of the
behavior of fig.~\ref{fig} and of  relation (\ref{osc1}) (or
(\ref{osc2})), i.e. making the scale factor  non-zero at the
bounce. However, in this work we desire to present the basic
characteristics of cyclic behavior in general collisionless
braneworld models, and thus we will not examine in detail the (in
any case non-complete) handling of the singularity. Our model
shares this disadvantage of all cyclic models. We will return to
this subject in the discussion section.

In the above investigation we have been restricted to stationary
solutions of the form (\ref{stationary}), case in which stable
time evolution is implied easily. The question is what can be said
about the full dynamics of equations (\ref{eomfull}) and
(\ref{constraints}), where $H_0^2$ and physical brane distance
$D(t)$ can be varying. Fortunately, solutions of the full dynamics
seem to consist of stationary ones and the transitions between
them \cite{brcod,Tetradis01}. Therefore our stationary
investigation is sufficient. We will return to this subject in the
next section.

\section{Discussion-Conclusions}\label{discussion}

In the aforementioned analysis we considered general braneworld
models characterized by the action (\ref{action}), the conformal
metric (\ref{metric}), and the general potentials (\ref{bulkpot})
and (\ref{branepot}). Performing both analytical and numerical
calculations we showed that the full $5D$ dynamics allows for
stationary solutions corresponding to oscillatory scale factor of
the physical brane and therefore to cyclic universes. In
statistical terms cyclicity corresponds to $\approx 4\%$ of the
physical solutions. Our investigation is completely $5D$, cyclic
behavior arises naturally and is induced on the brane by the full
dynamics, and it is not a result of a modified $4D$ dynamics, with
fine-tuned parameters or specific assumptions in the Friedmann
equation. Furthermore, we do not use an explicit brane state
equation, considering just the bulk scalar field (the decays and
interactions of which will eventually fill the physical brane with
the conventional content \cite{apostol}). As we mentioned in the
introduction this full $5D$ approach is necessary in order to
confront the arguments of the authors of \cite{Linde01}. Indeed,
their allegations that one cannot transit to an effective $4D$
theory (integrating the action over $y$), solve the equations
there and then return naively to the $5D$ description (adding
time-dependence by hand), are correct. Doing so, the results are
not self-consistent (especially the boundary conditions are not
satisfied) and the authors of \cite{Linde01} use this fact as a
central argument against the cyclic scenario. However, our
consistent $5D$ analysis reveals that cyclic behavior is possible.

Another important feature of the present study is that cyclic
universes do not require brane collisions. Thus, we avoid the
known problems concerning such a description, which force
ekpyrotic model to successively more complicated versions. On the
contrary, the branes do not move at all and the system is stable
(stationary solutions are a stable fixed point
\cite{brcod,Tetradis01}). Furthermore, in our model, expansion and
contraction take place in the 3+1 branes, and in all 3+1 slices in
general, while the fifth dimension remains unaffected. The $4$
spacial dimensions shrink periodically to an $1D$ string and
re-expand. This is in a radical contrast with the cyclic models
with extra dimensions, where the extra dimension is the one that
gets contracted (the fifth in \cite{ekpyrotic1} or the eleventh in
\cite{Khoury.rebound}). Cyclicity seems to re-obtain its
``physical" meaning.

Our $5D$ investigation is general and does not involve extra
assumptions, fine-tunings or specific potential forms. We result
to periodic, cyclic, homogenous and isotropic universes, where the
scale factor changes smoothly from expanding to contracting. An
observer on the physical brane feels successively accelerated
expansion, decelerated expansion, turnabout, accelerated
contraction, decelerated contraction, bounce e.t.c, and a
promising signature of the cyclic behavior would be the measure of
the varying rate of the Hubble constant. The cycles period, given
in (\ref{period}), can be arbitrary, depending on $B(0)$, i.e. on
the value of the warp factor on the physical brane ($\theta$ is
bounded from above and therefore cannot act as a period-decreasing
factor). A very interesting conclusion comes from the insertion of
observational results in our model, which was not made above in
order to remain as general as possible. Explicitly, if we use the
fact that the cosmological constant of our Universe is extremely
small ($\approx{\cal O}(10^{-47})\ \text{GeV}^4$), and assuming a
reasonable $M_5$ value of ${\cal O}(10^{19})$ GeV, the first two
boundary conditions in relation (\ref{junctions}) provide in
general a huge value for $e^{B(0)}$ ($\approx{\cal O}(10^{45})$).
This is in consistency with the scaling transformation of
\cite{brcod}, which allows us, in a solution, to scale the
parameters by $e^{-S}$ and add to the warp factor the constant
$S$, and acquire another solution. Therefore, the extremely small
cosmological constant of the observable universe leads the
cyclicity period to be around $T\approx{\cal O}(10^{13})$ years
and the maximum scale factor value, given by (\ref{minmax}), to be
$a_{max}\approx {\cal O}(10^{28})$ m (where the decimal exponents
in these rough estimations can vary by 1 or 2, depending on
$B'(0)$ and $\theta$ values). Luckily enough, the smallness of the
cosmological constant excludes oscillatory models with small
periods in astronomical terms. In more foundational words, the
reason that made the cosmological constant that small, is the same
that makes the cycle period and the size of the Universe that
large.

In this work we have been restricted to stationary solutions,
where the subclass of them that possesses $H^2<0$ corresponds to
eternal cyclic behavior with constant period. Numerical
investigation of the full dynamics seem to consist of such
stationary solutions and the transitions between them
\cite{brcod,Tetradis01}. In such transitions $H_0^2$ on the
physical brane can chance sign, leading to a form of ``chaotic
cyclicity", where large intervals of (non-periodic in general)
oscillatory behavior could be followed by large intervals of
conventional evolution and vice versa. In this case, an initial
Big Bang and/or a final Big Rip or Big Crunch (in conventional
terms) could be possible. Another interesting possibility would be
the exploration of our model with cosmological constants being
piecewise constant functions of time, reflecting cosmological
phase transitions, which could also lead to chaotic cyclicity.
Note however that numerical confirmation of such behaviors is very
hard due to the small probability of cyclic stationary solutions
($\approx 10^{-2}\%$ as we have already mentioned). These subjects
are under investigation.

In order for a model to serve as a description of nature, it has
to explain the basic physical key issues. Especially for cyclic
cosmology, amongst others these are the entropy evolution and,
probably the most pressing issue, that of a fuller understanding
of the bounce and the handling of the singularity. Our model
provides a consistent background for cyclicity and it reveals how
such a behavior arises from the full $5D$ dynamics. However, since
braneworlds and brane cosmology in general arise as limits of a
multi-dimensional theory unknown up to now, the $5D$ results have
a phenomenological character and must be considered from this
point of view. Definitely, a complete explanation and
apprehension, and a successful confrontation of the aforementioned
subjects, can only come through a higher-dimensional, fundamental
theory of nature. For the moment we have to rely on the relevant
research on cyclic cosmology, linearized gravity, M-theory and
strings, which has improved our knowledge on these issues. These
results can be embodied in our analysis. The most hopeful effort
is the use of quantum fluctuations in order to tame the
singularity, which effectively is translated into a modification
of gravity by the scalar field
\cite{Bojowald01,ekpyrotic3,ekpyrotic4}. Alternatively, using loop
quantum gravity we could modify non-perturbatively the dynamical
equations leading to a singularity resolution as in
\cite{Maartens}. Concerning the entropy, we could include the
relevant discussion in our investigation. The argument of the
authors of \cite{Turok.sing,Turok.simplified} about maximum amount
of entropy possible in de Sitter spacetime, may lead our model to
have a maximum cycle number between $10^{20}$ and $10^{30}$.
However, the idea of the causal patch \cite{Turok.sing} is
probably the best way of handling the entropy problem so forth,
and there are some interesting recent works on the subject which
give a boost on cyclic cosmology \cite{entropy}.

Let us close this discussion section with some comments on the
role of the brane tensions and of the bulk cosmological constant
in our model. As can be numerically confirmed, setting them to
zero makes it almost impossible to satisfy the boundary conditions
obtaining $H^2<0$ and singularity absence in [0,1] (this can be
achieved only through a careful fine-tuning since our random
choice procedure gives an one-digit number of such solutions in
$10^6$ parameter multiplets). On the other hand, as we showed in
\ref{B}, in the case where $\Lambda$, $\lambda_0$ and $\lambda_1$
are the only non-zero parameters, an $\approx10^{-1}\%$ of the
random parameter choices, or $\approx6\%$ of the solutions,
correspond to $H^2<0$. In mathematical terms, $\Lambda$,
$\lambda_0$ and $\lambda_1$ are requisite in order to acquire a
solution with $H^2<0$ in the full dynamics, in a natural and not
in a fine-tuning way. In terms of physics, it is the dark energy
that lies in the background of the oscillatory mechanism and
allows for cyclicity to realize. Adding the fact that it
determines the cycles period and the maximum scale factor value,
we conclude that dark energy is crucial in the described model.
This brings it closer to the ekpyrotic paradigm of the literature.

In this work we examine general braneworld models and we show that
cyclic behavior can naturally arise from the full $5D$ dynamics.
One important feature is that brane collisions are not required,
on the contrary the branes remain stable, and the cyclicity takes
place on the $4D$ geometry not on the extra dimension. Another
significant result is that the smallness of the cosmological
constant of the observable universe pushes the cyclic period and
the scale factor to astronomical large values, an essential
requirement for the establishment of cyclic cosmology as a
realistic alternative paradigm. Furthermore, we indicate the
possibility of a ``chaotic cyclicity", that is extremely large,
non-periodic, cyclic intervals followed by extremely large
intervals of conventional evolution and vice versa. After these,
the model shares both the
advantages and disadvantages of cyclic cosmology.\\

\paragraph*{{\bf{Acknowledgements:}}}
The author  acknowledges partial financial support through the
research program ``Pythagoras'' of the EPEAEK II (European Union
and the Greek Ministry of Education).

\end{document}